\documentstyle[12pt,aasms4]{article}

\slugcomment{}

\lefthead{Kuhn, Potter, and Parise}
\righthead{Imaging Polarimetric Detection ...}

\begin{document}

\title{Imaging Polarimetric Observations of a New Circumstellar Disk System}

\author{J.R. Kuhn}

\and

\author{D. Potter}

\and

\author{ B. Parise\altaffilmark{1}}

\affil{ Institute for Astronomy University of Hawaii Honolulu-HI-96822}
 
\altaffiltext{1}{Ecole Normale Superieure, Paris, France}



\begin{abstract}
Few circumstellar disks have been directly observed. Here we use
sensitive differential polarimetric techniques  to overcome atmospheric
speckle noise in order to image the circumstellar material around HD
169142. The detected envelope or disk is considerably smaller than
expectations based on the measured strength of the far-IR excess from
this system.
\end{abstract}




%

\section{Introduction}

Progress toward understanding the evolution of disk-like stellar
systems will be greatly aided by improved techniques and new imaging
measurements of candidate objects. Photometry from the UV-to-far IR of
several likely unevolved stars show spectral evidence of circumstellar
material (cf. Malfait et al. 1998). The problem of how  
these will evolve toward planetary systems is of great interest, but we
are far from understanding the general conditions which lead to planet
formation from premainsequence objects. With more than 50 planetary
systems detected (Marcy and Butler, 2000) it is the observational clues
to the premainsequence systems which are particularly scarce -- only a
few resolved observations of circumstellar disk systems exist.

Observing faint circumstellar material environments from ground-based
telescopes is typically a feat for adaptive optics (AO) but it is more
aptly described as a dynamic range problem, something which is not
automatically assured with current AO systems. In particular we show
here how a 4m telescope at a good site, operating in the H band, with
only low order (tip-tilt) wavefront compensation and high dynamic range
detectors can be adequate for extracting smaller-than 1\% circumstellar
signals within an arcsecond of a bright star. The dusty environment of
our target (HD169142) was unsuccessfully searched for earlier (Harvey
et al.  1996).

The general problem of extracting a faint optical/IR signal from the
scattered light due to atmospheric seeing and telescope diffraction
requires a stable point spread function (PSF) and a high dynamic range
measurement. High dynamic range can be effectively achieved with
coronagraphic observing techniques (assuming high Strehl ratio is
achieved) and using fast framing IR array detectors but, unfortunately,
from any ground-based optical system the stability of the point
response function is limited by atmospheric speckle noise.
Differential, multi-wavelength, observations (Racine et al. 1999) have
been suggested as a method to minimize such noise, although their
approach may be limited by the chromaticity of the residual noise (but
see Marois et al. 2000 for a discussion of this technique). Here we
demonstrate how differential polarimetric techniques can have
significant detection threshold advantages. Mauron and Dole (1998)
have achieved excellent detection thresholds using a similar dual
beam polarization
technique with non-imaging detectors.

To illustrate this we consider the problem of detecting the linearly
polarized scattered light from a circumstellar disk in the presence of
unpolarized scattered light from the central star.  We assume a
detector which simultaneously images two orthogonal polarization states
on distinct regions of the IR camera array. For small scattering angles
(less than a few arcseconds) the atmospherically scattered speckle
patterns in each polarized brightness component are indistinguishable,
so that the difference image constructed from orthogonal polarizations
removes the scattered light, leaving a polarized brightness
circumstellar image. For real non-photon counting detectors the
residual noise in the difference image will be limited by calibration
flat-fielding uncertainties. These can be minimized with a
``double-chop'' technique which exchanges the orthogonal polarization
illumination on the array at the expense of introducing some speckle
noise into the measurement.  We describe the post-flat-field residual
pixel-to-pixel gain variations in the two images by $(1+\epsilon_i(\vec
r))$ where i=1,2 and the $\epsilon_i(\vec r)$ are two different
functions which account for the residual flat-field errors in the two
image sections. Using least-squares algorithms described by Kuhn et al.
(1991) we expect $\epsilon_i$ to be of order $10^{-3}$.

We take
$I(\vec r)$ to represent the scattered light scene in one subimage snapshot
and $I'(\vec r)=I(\vec r)(1+\delta(\vec r))$ to express the image scene (including
speckle changes) in a second snapshot a short time later. Here $\delta(\vec r)$ describes the
amplitude of the speckle variation from one snapshot to the next. We
take $p_{\pm}(\vec r)$ as the contribution to the polarized image flux from the
circumstellar material.
The effect of two polarizers (labeled + and -) is to produce an image
scene of scattered (proportional to the input) and circumstellar light, i.e.
$K_+I(\vec r)$ and   $p_+(\vec r)$. The data recorded
for the + analyzer at the camera is $d_+=(1+\epsilon_1)(K_+I+p_+)$ and
for the - state 
$d_-=(1+\epsilon_2)(K_-I+p_-)$. A second snapshot obtained with the
+ and - polarization images interchanged on the detector yields $d'_+=(1+
\epsilon_2)(K_+I'+p_+)$ and $d'_-=(1+\epsilon_1)(K_-I'+p_-)$. By combining
the sum of the differences we obtain 
\begin{eqnarray}
((d_+-d_-)+(d'_+-d'_-))/I(\vec r)~=~
2(p_+(\vec r)-p_-(\vec r))/I(\vec r) + 2(K_+-K_-)  + \nonumber \\
(\epsilon_2(\vec r)+\epsilon_1(\vec r))(K_+-K_-) + 
(K_+-K_-)\delta (\vec r)  +
(\epsilon_1(\vec r)-\epsilon_2(\vec r))(p_+(\vec r)-p_-(\vec r))/I(\vec r)   
\end{eqnarray}
And we've ignored terms which
are as small as the product of $\epsilon$ and $\delta$. For most of
these observations the detector read noise is insignificant.

Isolating the circumstellar polarization signal, the first term on the
rhs of (1), is the aim of this analysis.  All the other terms on the
rhs of (1) describe the effects of flat-fielding and speckle noise.
During observing we will accumulate temporally interleaved images
corresponding to $d_+$, $d_-$, $d'_+$, and $d'_-$.  The polarized flux
constants, $K_+$ and $K_-$, can be empirically obtained from virtually
all of the illuminated pixels by scaling the subimages to minimize
their difference. Since there are many pixels $K_+$ and $K_-$ can be
determined to an accuracy of better than $10^{-4}$ (significantly
better than the photon noise in a single pixel). In this case it is
clear that the noise in the polarization due to flat-fielding errors
(the third term on the right of this equation) can be minimized. By
interleaving snapshots and averaging many frames we can also force
$\delta (\vec r)$ to be less than $10^{-2}$ so that the fourth term of
(1) is also negligible. The double-chopping polarization analysis
should eliminate speckle noise from our measurements to the level of
$10^{-3}$ of the background as long as we accumulate at least $10^6$
background photons per pixel to overcome photon noise (and read noise),
and as long as we can minimize the flat-fielding errors, $\epsilon$.
Flat-fielding uncertainties have been the largest concern in
applying these techniques.

\section{Observations}

The United Kingdom Infrared Telescope (UKIRT) and its facility IR
polarimeter are well suited for this program. The observations
described here were obtained Sept. 1-3, 2000, with the UKIRT IRCAM InSb
camera and the IRPOL polarimeter which provides simultaneous e- and
o-ray images through a ``cold'' Wollaston prism. In the H-band the image
deviation (and field-of-view) is about 5 arcseconds and the camera
scale is 0.0814 arcseconds per pixel. A warm rotating waveplate is
mounted ahead of the image plane mask to allow orthogonal polarization
states to be interchanged on the detector.

Approximately 10 candidate disk systems from Mannings and Barlow (1998)
and Coulson et al. (1998) were selected. Typical H band observations of
these 7-11 magnitude sources were obtained by coadding between 10-100
0.1-1 sec exposure images. The waveplate was rotated by 22.5 degrees
(for a total of 4 waveangles) and exposures were repeated. The
telescope was also offset by as much 0.3 arcsec to 3 spatially
displaced pointings and the polarization sequence was repeated.
Averaged data from the displaced images helped to reduce residual
flat-fielding noise.  Unpolarized and polarized standard stars from the
IRPOL database were also observed.

Several different techniques were used to obtain flat-fielding data.
The best calibration was obtained from extended object planetary
images. A set of 8 displaced Uranus images were obtained for each of
the 4 waveplate angles. The image displacements ranged up to 1.5
arcseconds and were chosen in non-redundant directions in order to
apply a least-squares flat-fielding algorithm (Kuhn et al. 1991).

All observations were obtained from similar coadded image sequences
using a common IRCAM/IRPOL optical configuration. To define
corresponding pixels (scale and offset) in the displaced and polarized
subimages we observed a collection of nearby point sources (the M15
globular cluster). Dark/bias corrections for all frames were derived
from equivalent integration time images obtained immediately before
each coadded sequence. Separate flat-fields were computed and applied
for data from each waveplate angle.

Several program stars are still being analyzed but one object shows
unambiguous evidence of a dusty circumstellar environment with the
polarization signature of a face-on disk. HD169142 (SAO 186777, IRAS
18213-2948) was observed Sept. 3, 2000 at 13:43 UT. A polarization sequence
was obtained using 30 2s exposures. A shorter sequence of 20 0.3s
exposures were also obtained to obtain the total stellar flux since
longer exposures are saturated in the image core. The seeing during the
elapsed 30min of observations was variable but approximately 0.5" at an
airmass of 1.75.  This is a bright,  $m_B$=8.4  B9Ve star at an
estimated distance of 145pc (Sylvester et al. 1996). It has been
observed by Yudin et al.  (1999) to be polarimetrically variable over
timescales of a day.  We are not aware of any spatially resolved
observations of this object although models of its spectral energy
distribution (Malfait et al.  1998) suggest that it has two extended
circumstellar shells or disks, with the outer one extending from a few
tenths to several arcsecond from the star. It should be noted that its
conventional identification as a young Herbig AeBe star is not ironclad
-- it shows no evidence of proximity to nebulosity or star forming
regions (Herbig 2001).

\section{Analysis and Discussion}

Figure 1 shows the mean intensity and Stokes Q and U polarization
images obtained from HD169142. The Q and U images are obtained as in
eq. (1) by differencing and averaging orthogonal polarization data. The
similar, but 45 degree rotated Q and U data, show a signature
of an optically thin scattering source surrounding the central star.
Figure 2 shows how the polarization direction is generally organized in
concentric rings around the central source. The total polarized flux
is only $7.5\times 10^{-3}$ of the total stellar flux. The infrared
excess to stellar luminosity ratio for this system is also small
($8.8\times 10^{-2}$, Yudin et al. 1999). 

The polarization fraction is not uniform, but shows clumpiness over the
scale of the image. The polarized brightness in Figure 3 (equal to the
polarization fraction times the mean H band surface brightness
intensity) provides evidence of scattering inhomogeneity within the
circumstellar cloud. The slight elongation of the polarized brightness to the
upper right and lower left in Figure 3 is also a robust
feature of the HD169142 circumstellar cloud. 

Although the dominant errors in these measurements are from the
residual flat-fielding calibration uncertainty, the detected
polarization within 1.7 arcsec of the star is highly significant.  We
estimate the noise from observations of an unpolarized source.
Figure 4 plots the ratio of the circularly averaged polarized flux
divided by the unpolarized brightness, from HD169142 and from an
unpolarized standard (dotted curve). The figure shows the expected
polarization fraction bias (Serkowski 1962) of about 0.005 in the
standard star observations (dotted line). These data have an RMS
polarization uncertainty of 0.0012. Thus, subject to the assumption
that the intrinsic circumstellar material polarization is a weak
function of distance from the star, we conclude that the drop in
polarized light fraction near 1.5 arcseconds is evidence of a disk
edge.  In any case, the non-zero polarized light fraction measurements
within 1.5 arcseconds are highly significant, while the light beyond
1.7 arcseconds from the source is consistent with unpolarized scattered
light from the telescope and atmosphere.  Apparently the disk in
HD169142 has a radius of about 1.5 arcseconds.  In contrast the thermal
model which Malfait et al.  (1998) required to describe this system
implied a much larger disk size, roughly an order of magnitude bigger
than what we observe here.

These observations reveal a faint circumstellar envelop around
HD169142. The relatively small angular extent of this disk-like cloud
has been measured without high-order adaptive optics. Its detection
here demonstrates the utility of dual-beam imaging IR polarimetry for
overcoming atmospheric speckle noise and for achieving high photometric
dynamic range in the angular vicinity  of bright sources.


\acknowledgments
The United Kingdom Infrared Telescope is operated by the Joint Astronomy
Center on behalf of the U.K. Particle Physics and Astronomy Research Council.
We thank the Department of Physical Sciences, University of Hertfordshire
for providing IRPOL2 for the UKIRT. We are indebted to Chris Davis, Sandy
Leggett and the UKIRT staff for their support during these
observations. We also acknowledge useful discussions with George Herbig and
Ingrid Mann concerning the nature of HD169142 and its dusty environment. 

\clearpage

\clearpage

\figcaption[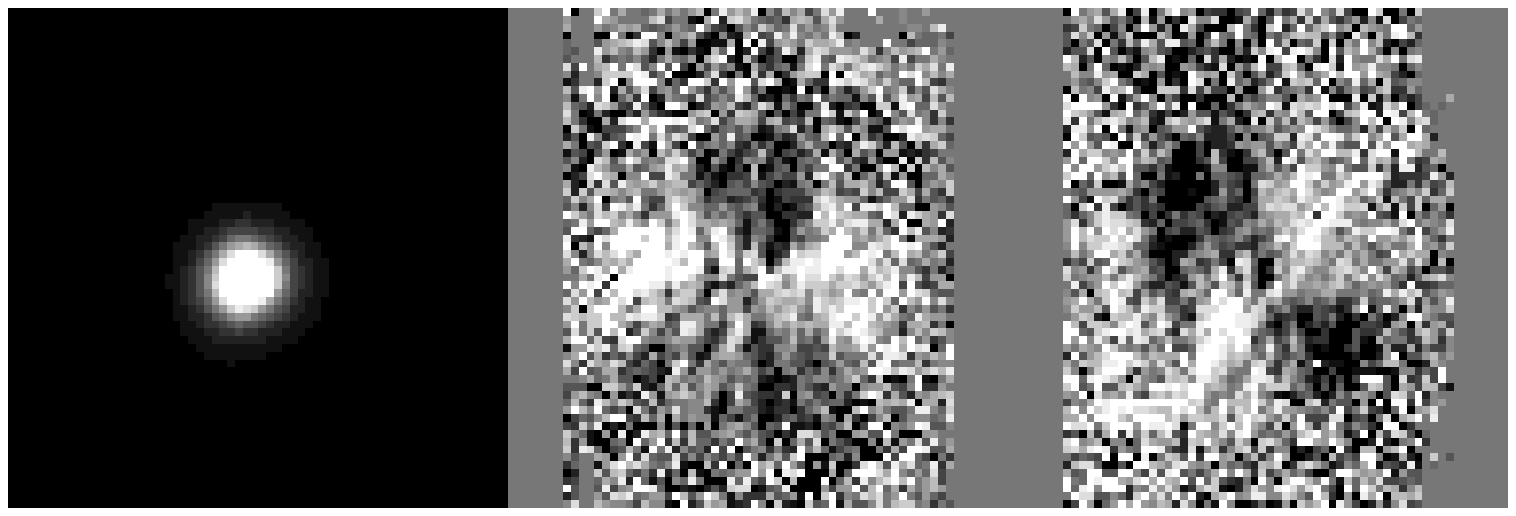]{ Average intensity, Stokes Q, and U
polarization images are shown here (normalized as the first term on the
right hand side of eq. (1)).  The full-width at half-maximum of the
intensity image is 7 pixels (0.57 arcseconds). The center panel shows
the vertical polarization while the right hand panel displays
polarization oriented from upper left to lower right as positive. The
full range from dark to light corresponds to -1.5 to 1.5 \% polarization.
The spatial scale of each image is 5.2 arcseconds.
\label{fig1}}

\figcaption[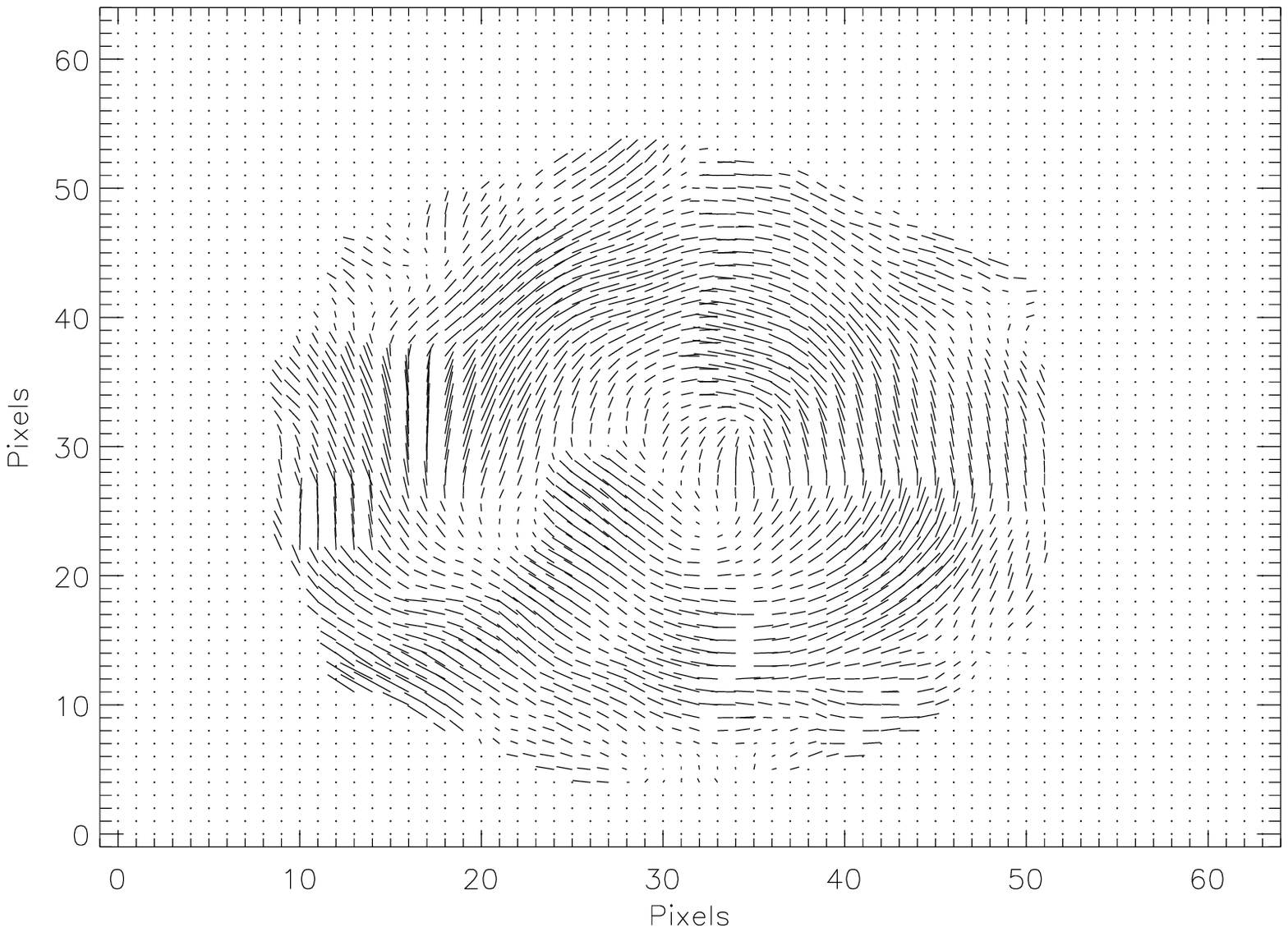]{The polarization derived from Q and U measurements
is plotted here. The length of the plotted line segments is proportional
to the polarization fraction (about 1.5\% in the strongest pixels). The
Q and U data were smoothed with a 5 pixel boxcar average before plotting and
a minimum intensity threshold defined pixels to be plotted.  \label{fig2}}

\figcaption[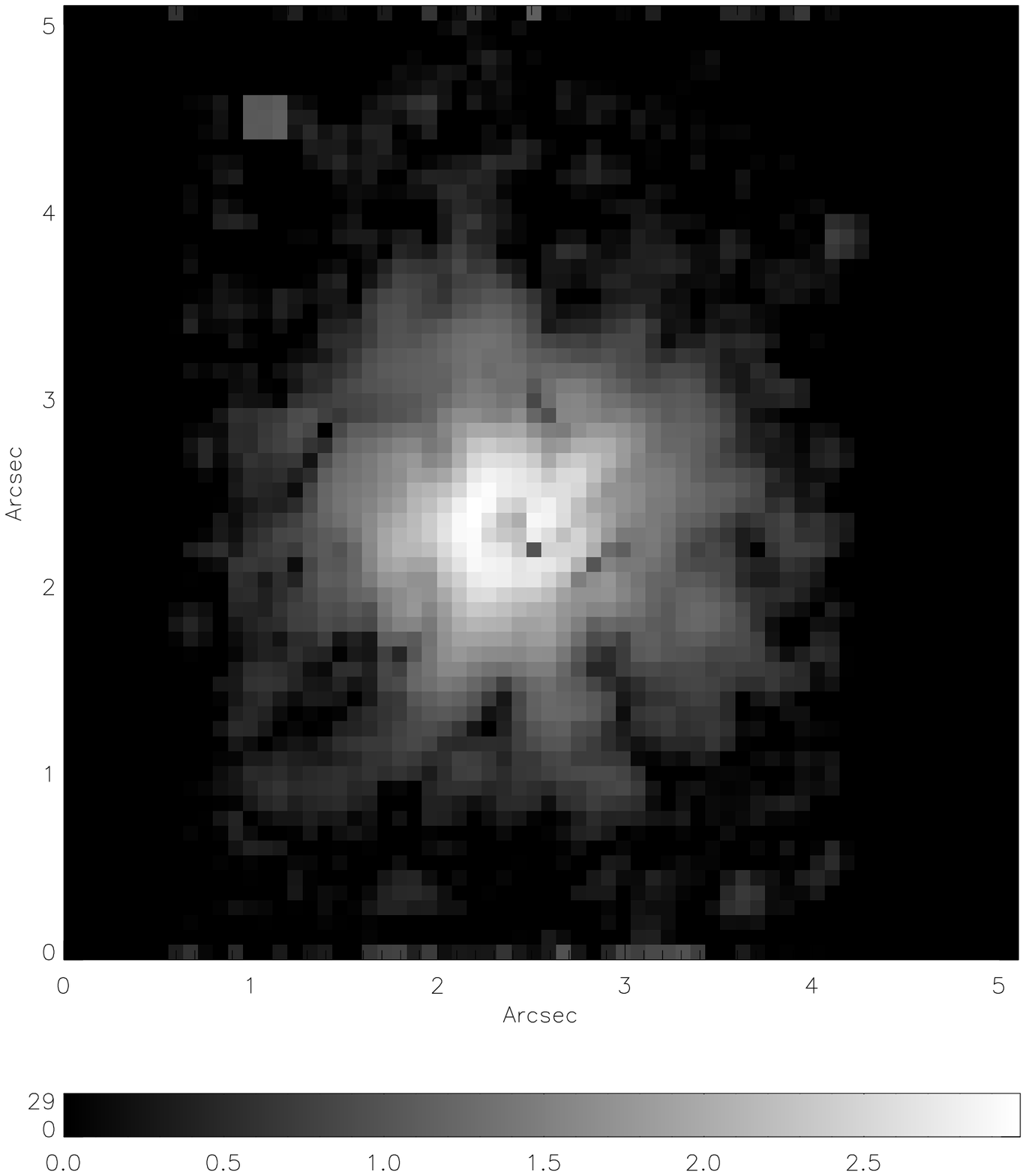]{The logarithm (base 10) of the polarized brightness is
displayed here. North is to the right and east is up in this
image. \label{fig3}}

\figcaption[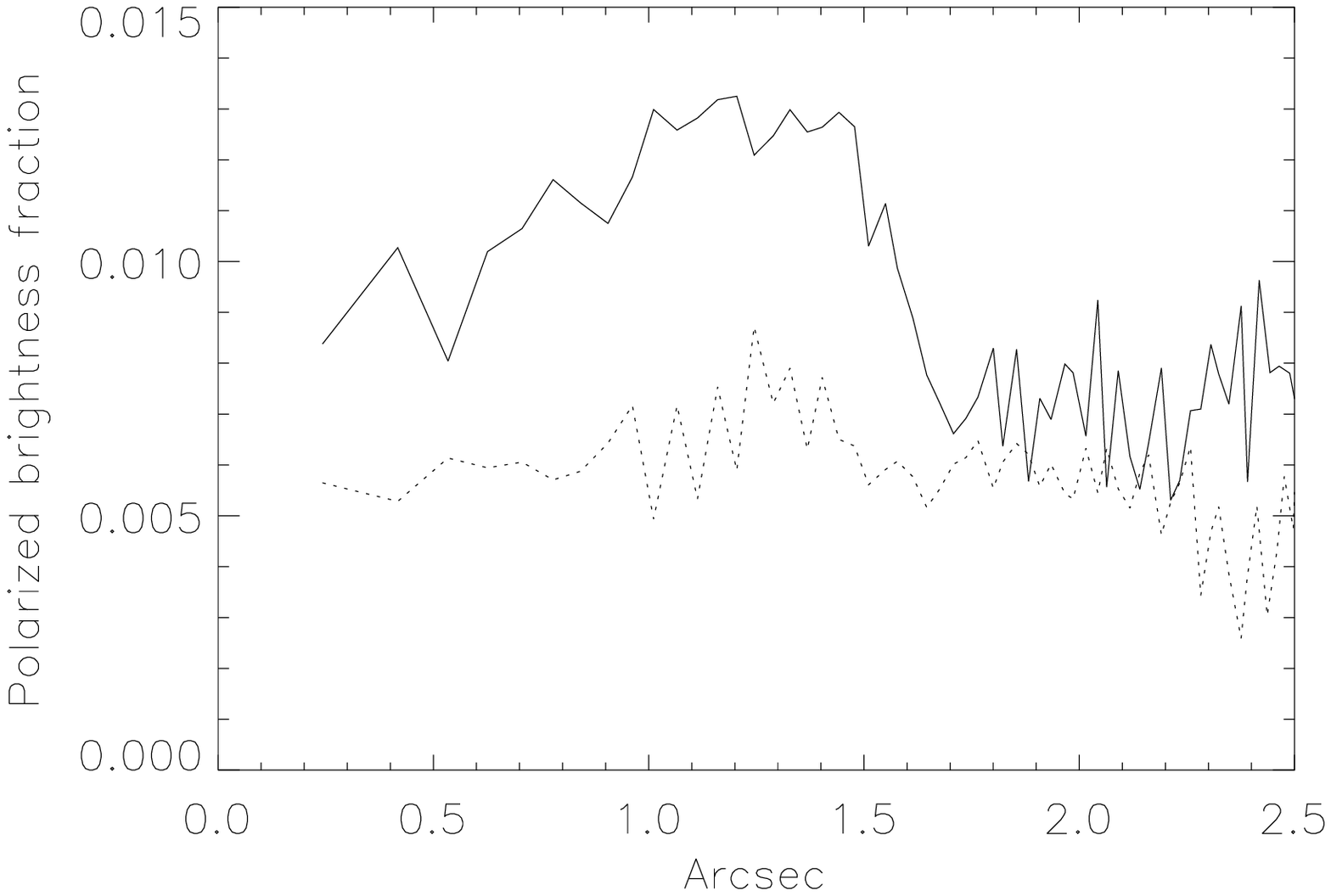]{The circular average of the polarized brightness normalized by the mean brightness 
as a function of distance from the central star is plotted here. The dotted line shows the noise level from a similar calculation for an unpolarized calibrator star. The circumstellar disk extends to a radius of approximately 1.5 arcseconds. \label{fig3}}


\clearpage

\plotone{figure1.ps}


\plotone{figure2.ps}


\plotone{figure3.ps}


\plotone{figure4.ps}


\begin{thebibliography}{}

\bibitem[Coulson et al. 1998]{coul98} Coulson, I.M., Walther, D.M., and
Dent, W.R.F. 1998, MNRAS, 296, 934.
\bibitem[Harvey et al. 1996]{harv96} Harvey, P.M. Smith, B.J., Difrancesco, J., Colome, C. 
and Low, F. 1996, \apj 471, 973.
\bibitem[Herbig 2001]{he01} G. Herbig, personal communication (2001).
\bibitem[Kuhn et al. 1991]{ku91} Kuhn, J.R., Lin, H., Lorenz, D. 1991, PASP,
103, 1097. 
\bibitem[Malfait et al. 1998]{mal98} Malfait, K., Bogaert, E., and 
Waelkens, C. 1998, AA, 331, 211.
\bibitem[Mannings and Barlow 1998]{man98} Mannings, V., Barlow, M.J. 1998
\bibitem[Marcey and Butler 00]{mar00} Marcy, G., Butler, R. P. 2000, PASP, 112, 137.
\bibitem[Marois  et al. 2000]{mar2000} Marois, C., Doyon, R., Racine, R, Nadeau, D. 2000, PASP, 112, 91.
\apj , ~497, 330.
\bibitem[Mauron and Dole]{mau98}Mauron, N., Dole, H. 1998 AA, 337, 808.
\bibitem[Racine et al. 1999]{rac99} Racine, R., Walker, G., Nadeau, D., Doyon, R., Marois, C. 1999, PASP, 111, 587.
\bibitem[Serkowski 1962]{ser62} Serkowski, N, 1962, Advances in Astronomy and Astrophysics, 1, 304.
\bibitem[Sylvester et al. 1996]{syl96} Sylvester, R. J., Skinner, C.J., Barlow,
M.J., and Mannings, V. 1996, MNRAS, 279, 915.
\bibitem[Yudin et al. 1999]{yud99} Yudin, R.V., Clarke, D., and Smith, R.A.
1999. AA, 345, 547.

\end{thebibliography}
\end{document}